\documentstyle[12pt,epsf,psfig]{article}
 
\def\be{\begin{equation}}
\def\bea{\begin{eqnarray}}
\def\ee{\end{equation}}
\def\eea{\end{eqnarray}}

\begin{document}
 
\title{Pair Connectedness and Shortest Path Scaling in Critical Percolation}
\author{Peter Grassberger\\
{\sl \small NIC, Forschungszentrum J\"ulich, D-52425 J\"ulich, Germany}}
\maketitle
 
\medskip
\medskip
 
\begin{abstract}
\noindent
We present high statistics data on the distribution of shortest path lengths 
between two near-by points on the same cluster at the percolation threshold. 
Our data are based on a new and very efficient algorithm. For $d=2$ they 
clearly disprove a recent conjecture by M. Porto {\it et al.}, Phys. Rev. {\bf E 58}, 
R5205 (1998). Our data also provide upper bounds on the probability that two 
near-by points are on different infinite clusters.

\end{abstract}

\section{Introduction}

Although percolation is a problem which has been studied in great detail during the 
last few decades \cite{stauffer}, and although many exact results are known by now, 
there are still open questions which either have not been studied at all, or which 
have not yet been understood. 

In the present note we study spreading of percolation in form of an epidemic process
\cite{grass82}. In the physics literature, this is often called Leath growth of clusters 
\cite{leith}. In this process (which we assume to proceed in discrete time steps) 
one starts with an infected seed as single `growth site', and keeps at each time a
list of growth sites. In the next time step, the list consists of all wettable 
sites which are nearest neighbors to any of the present growth sites, while the 
old list is canceled. A site is wettable if it had not been a growth site before, 
and if it can be occupied (in site percolation) or it is connected to the present 
growth site by an occupied bond (in bond percolation). It is well known that the 
distribution of growth sites satisfies at the critical point and for large times $t$ 
the scaling law \cite{grass82,grass92}
\be
   \rho({\bf x},t) = {1\over t^{1+2\beta/\nu_t}} \phi(r/t^z)\;.
\ee
Here, $\beta$ is the well known critical exponent governing the fraction of sites 
occupied by the infinite cluster \cite{stauffer}, $\nu_t$ is the critical 
exponent governing the correlation time if one goes off-critical, and $z=\nu/\nu_t$
is a dynamical exponent. The inverse of the latter is often called $d_{\rm min}$. 
Finally, $\phi(\zeta)$ is a universal scaling function. In a purely geometrical 
interpretation of eq.(1), $t$ is often called the `chemical distance' between 
the site ${\bf x}$ and the origin, as it counts the number of lattice steps 
{\it on the cluster} needed to reach ${\bf x}$ from the origin (i.e. the length 
of the shortest `chemical path').

The main problem we want to study is the behavior of the scaling function $\phi(\zeta)$ 
for small $\zeta$. This has been studied recently by Porto {\it et al.} \cite{porto} 
(see also \cite{dokhlo}). As usual one expects a power law, 
\be
   \phi(\zeta) \sim \zeta^{g_1}\;,\quad \zeta \to 0 \;.     \label{phi}
\ee
The authors of \cite{porto} used an analogy with self avoiding random walks (SAW) to 
conjecture 
\be
   g_1 = d_{\rm min} - \beta/\nu  \qquad ({\rm conjectured})               \label{g1}
\ee
where $d$ is the dimension of the lattice in which the cluster is embedded. 
This conjecture was then checked numerically and found to be satisfied 
\footnote{In the present paper, $\rho({\bf x},t)$ is the density averaged over {\it all}
clusters containing the origin, not only over the infinite one. It seems that the 
authors of \cite{porto} wanted their argument to apply to the incipient infinite 
cluster only. They implemented this by analysing only clusters with chemical radius 
$t_{max}>2000$, from which they presented results for chemical distances up to 1800. 
If there were a substantial dependence on the chemical radius, this would lead to 
spurious violations of scaling since $t_{max}$ introduces a new scale. But this 
dependence seems to be weak, as also assumed in \cite{dokhlo}.}

There are however a number of problems associated with that analysis. The first 
is that the analogy with SAW is not very stringent. The analogous equation to 
eq.(\ref{g1}) had been derived by deGennes \cite{degennes} using the fact that 
the end point density is the order parameter field in the SAW problem. Therefore, 
there is no anomalous critical exponent for the number of self avoiding loops 
(which have no end points), and this then gives an expression for the probability 
that a SAW nearly forms a loop which formally seems to involve mean field 
exponents. The authors of \cite{porto} essentially used these `mean field' 
arguments, although there is no analogous underlying field theory (the field 
theory for percolation is Potts-like \cite{fortuin}), and we see no logical basis 
for eq.(\ref{g1}).

\begin{figure}
  \centerline{\psfig{file=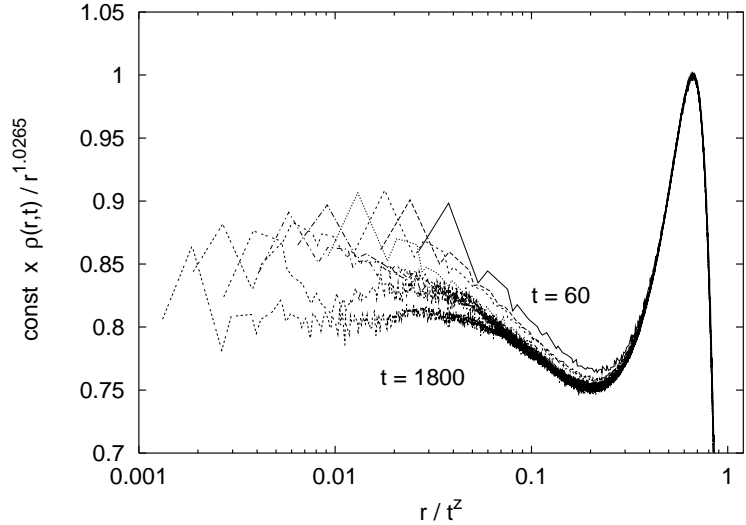,width=10cm,angle=270}}
\caption{\small  
Radial density of sites infected at time $t$, with 
arbitrary normalization, plotted against $r/t^z$ with $z=0.8844$ (corresponding to 
$d_{\rm min} = 1.1307 \pm 0.0004$; \cite{spread}). In order to enhance the significance 
of the plot, each curve corresponds to times $\in ]0.9t,t]$, and the data are 
divided by $r^{d_{\rm min}-\beta/\nu}$. The curves correspond to $t=60, 100, 140, 
200, 300, 500, 800, 1200$ and 1800.
}
\end{figure}

Similarly, the simulations used in \cite{porto} seem far from conclusive. In these 
simulations, clusters have been studied with $t\le 1800$ in $d=2$ and with $t\le 
800$ in $d=3$. In both cases, `more than 100,000' clusters have been analyzed, at 
several fixed values of $t$. 
In figure 1 we show results for $d=2$ obtained by their method. Each curve is 
based on more than $3\times 10^6$ clusters. In addition, in order to enhance 
the statistics, we have lumped together data with $0.9 t < t' \le t$ in each curve. 
In this way we arrive at statistics at least $10^3$ larger than those of 
\cite{porto}. In order to see more details we do not show simple scaling laws as 
in \cite{porto}, but show data multiplied by a suitable power of $r/t^z$. If 
eq.(\ref{g1}) holds, we expect to see a horizontal line for small x-values. 
Although our data are certainly not in contradiction with this, the very large 
statistical and systematic deviations from such a line render any precise 
statement impossible.

If we want a significant test of eq.(\ref{g1}), we have thus to proceed 
differently. Indeed, for fixed $r$ there is a much faster numerical 
method. While the above simulations need a CPU time $\propto t^{1.6}$ in order 
to analyze one cluster, our improved method is faster by roughly one power of 
$t$. 

Let us assume we want to test whether two sites ${\bf x}$ and ${\bf y}$ on a 
regular (hypercubic) lattice are on 
the same cluster, and want to measure the length of the shortest connecting path
if they are. Let us assume furthermore that the distance between ${\bf x}$ and ${\bf y}$, 
measured as the sum of the coordinate distances, is even (the case of odd distances 
will be discussed below).
Instead of growing a single cluster from a seed at either ${\bf x}$ or ${\bf y}$, 
we grow two clusters {\it simultaneously}, one starting from ${\bf x}$ and 
the other from ${\bf y}$. We stop the growth in any of the following cases:

(1) Both clusters have a common growth site. If this happen at time $t$, then there 
exists a path of length $2t$ passing through this growth site and connecting 
${\bf x}$ with ${\bf y}$. Since this is the shortest such path (otherwise the growth 
would have stopped before), in this case the chemical distance between 
${\bf x}$ and ${\bf y}$ is $2t$.

(2) The cluster growing from ${\bf x}$ dies (i.e., has no more growth sites). 
In that case, there cannot be a path 
from ${\bf x}$ which is long enough to reach ${\bf y}$, and ${\bf x}$ and ${\bf y}$ 
are on different clusters. Notice that both clusters cannot overlap. They cannot 
ovelap in a site which has the same chemical distance from both seeds, because 
they would then have been killed by rule (1) above. And they cannot overlap in a 
site which, say, has a smaller chemical distance from ${\bf x}$ than from ${\bf y}$. 
Such a site would first have been wetted by ${\bf x}$. In order to be wetted also 
from ${\bf y}$, it must have at least one other wettable neighbor which was not 
wetted, however, at the time step following its first wetting. This is not possible.

(3) The cluster growing from ${\bf y}$ dies. 

(4) $t$ reaches some upper bound $T$ specified at the beginning. In this case 
${\bf x}$ and ${\bf y}$ are either on different clusters, or the shortest path has 
length $>2T$.

If the distance between ${\bf x}$ and ${\bf y}$ is odd, we have to replace 
(1) by:

(1') A grow site of the first cluster at time $t$ coincides with a growth site of 
the other cluster at time $t-1$. In this case, the chemical distance is $2t-1$.

Let us denote by $p(t)$ the probability that an event dies because of rule (1). 
It is equal to the probability that two sites at a distance $r=|{\bf x}-{\bf y}|$ have 
chemical distance $2t$, which according to eqs.(1-3) is given by 
\be 
   p(t) = {\rho({\bf x}-{\bf y},2t)\over \sum_{t'=0}^\infty \rho({\bf x}-{\bf y},t')} \sim t^{-\lambda}
\ee
with
\be 
   \lambda = 1+{2\beta\over \nu_t} + z g_1\;.
\ee
Inserting here the conjectured eq.(\ref{g1}), we arrive at 
\be
   \lambda = 2 + {\beta\over \nu_t} \;. \qquad ({\rm conjectured})        \label{lambda}
\ee
   
During these simulations, we not only collect a histogram $n(t)$ of the times 
$t$ when rule (1) applies (which gives the distribution $p(t)$ when normalized), but 
also a histogram of times when the growth stops due to {\it any} reason or, 
equivalently, a histogram $N(t)$ indicating how many events have survived at 
least $t$ time steps. The numer $N(T)$ of events surviving until the very end 
gives an upper bound on the probability that ${\bf x}$ and ${\bf y}$ are on 
different {\it infinite} clusters, $P_{\rm diff} < N(T)/N$. If $P_{\rm diff}=0$ 
(which we expect for $d<6$; there are several infinite clusters for 
any $d\ge 2$, but the chance that they come close to each other should be zero), 
we expect $N(t)$ also to decay with a power law,
\be 
   N(t) \sim t^{-\mu}
\ee
with exponent $\mu \le \lambda-1$. This should be compared to the power with which 
single clusters survive, $-\beta/\nu_t$ \cite{grass82,grass92}. Thus according to 
the conjecture the difference in the powers is exactly 1, suggesting that the present 
algorithm is faster by one power of $t$.

\begin{figure}
  \centerline{\psfig{file=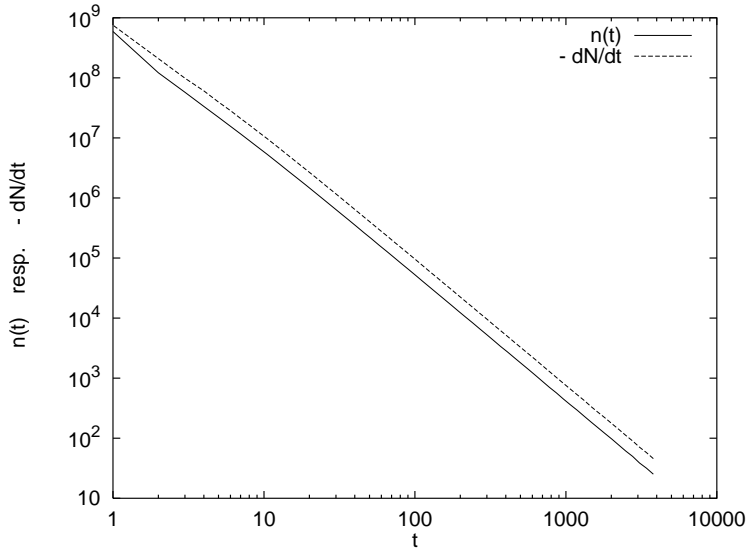,width=10cm,angle=270}}
\caption{\small  
Histograms of $n(t)$ and $-dN(t)/dt$ for $d=2$ bond percolation with ${\bf x}-{\bf y} = (1,1)$.
In order to reduce statistical fluctuations, data are averaged over intervals $[0.98 t,t]$.
}
\end{figure}

\begin{figure}
  \centerline{\psfig{file=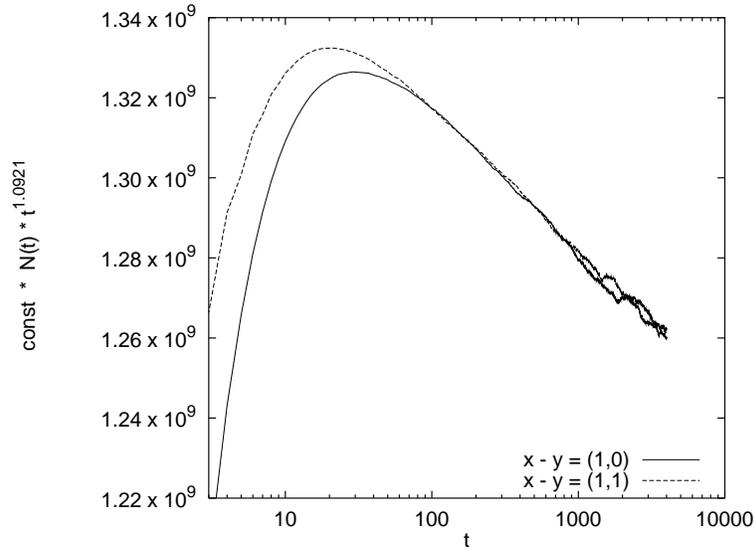,width=10cm,angle=270}}
\caption{\small 
Plot of $N(t) t^{1.0921}$ for $d=2$ with ${\bf x}-{\bf y} = (1,0)$ (lower curve) and 
for ${\bf x}-{\bf y} = (1,1)$ (upper curve). The numbers of configurations were chosen 
such that both curves coincide for large $t$. The exponent $1.0921 = 1+\beta/\nu_t$ 
is such that the curves should become flat for large $t$, according to the conjecture 
of \cite{porto}. Relative statistical errors are $\approx 1/\sqrt{N(t)}$. For $t=4000$, 
they are $\approx 2.6\times 10^{-3}$.
}
\end{figure}

\begin{figure}
  \centerline{\psfig{file=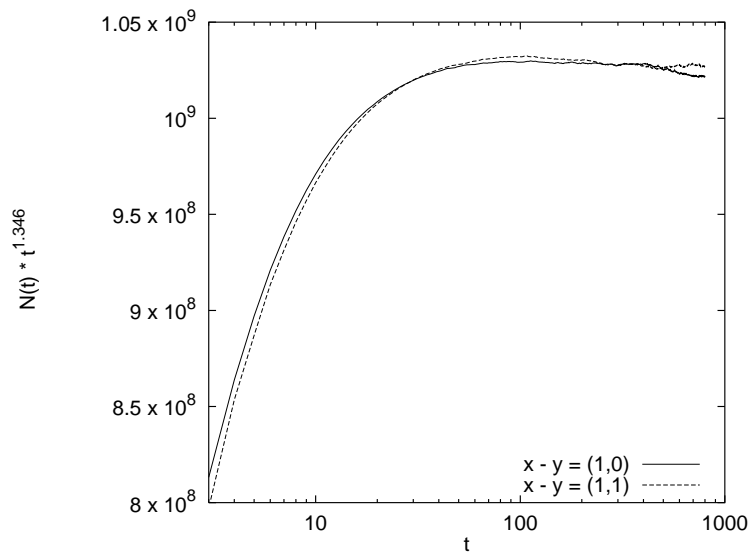,width=10cm,angle=270}}
\caption{\small
Similar to fig.3, but for $d=3$. This time we used $\beta/\nu_t = 0.3467\pm 0.0028$ from 
\cite{grass92,lorenz}. Relative errors for $N(t=800)$ are $\approx 3\times 10^{-3}$.
}
\end{figure}

In fig.2 we show, for $d=2$ and ${\bf x}-{\bf y} = (1,1)$, both $n(t)$ and $|dN(t)/dt|$.
We used bond percolation where $p_c=1/2$ exactly.
We see large deviations from power laws, but these deviations are the same for both 
curves. Obviously, if a pair of clusters dies, the chance that it dies because of rule 
(1) is finite and tends to a constant, and therefore
\be
   \mu = \lambda -1 \;.
\ee
If we accept this, we can obtain the most precise estimate of $\lambda$ from $N(t)$. 
In any case, even if this is not correct, we obtain from $N(t)$ a lower bound on 
$\lambda$. Since we shall see that this lower bound is larger than the value 
conjectured by \cite{porto}, this is sufficient to exclude the conjecture.

Results for $N(t)$ for $d=2$ are shown in fig.3, for ${\bf x}-{\bf y} = (1,1)$ and for 
${\bf x}-{\bf y} = (1,0)$. Each curve is based on $>10^9$ runs with $T=4000$, 
and took about 90 h CPU time on a fast workstation. In order to compare with 
eq.(\ref{lambda}), we multiplied $N(t)$ by $t^{1+{\beta\over \nu_t}}$, using the exact 
value $\beta =5/36$ \cite{stauffer} and the estimate $\nu_t = 1.5075 \pm 0.0004$
(corresponding to $d_{\rm min}=1.1306\pm 0.0003$). This estimate is based on new
simulations which follow exactly the lines of \cite{spread}, but have four times 
higher statistics. It is fully compatible with \cite{spread}. We see clearly that 
eq.(\ref{lambda}) is wrong since it gives a too small value. A precise estimate 
of $\lambda$ is hampered by the very strong corrections to scaling visible in fig.3. 
Fitting these corrections by terms $\sim 1/t$ relative to the leading term and 
assuming that $n(t)\propto N(t)$, we 
arrive at 
\be
   \mu = 1.1055 \pm 0.0010\;, \qquad \lambda = 2.1055 \pm 0.0010\;, 
        \qquad g_1 = 1.041 \pm 0.001.                      \label{mu}
\ee
Basing the analysis on $n(t)$ instead of $N(t)$, we would get $
\lambda = 2.105 \pm 0.002, g_1 = 1.041 \pm 0.003$.

A similar analysis for $d=3$ is shown in fig.4. This time $T=800$. For $p_c$ we 
used the value $0.2488126$ of \cite{lorenz}. The most precise value of $\beta/\nu_t$ 
can be obtained either by combining the value of $z$ from \cite{grass92} with the 
value of $\tau$ from \cite{lorenz}, or by using directly the estimate of $\beta/\nu_t$
from \cite{grass92}. The first gives $\beta/\nu_t = 0.347\pm 0.004$, the latter 
$\beta/\nu_t = 0.345\pm 0.004$. In addition we performed further simulations using the 
method of \cite{grass92}. Together, all these combine to our final estimate
$\beta/\nu_t = 0.3467\pm 0.0028$. As seen from fig.4, the corrections to 
scaling are even worse than for $d=2$, and the disagreement with eq.(\ref{lambda})
is much less pronounced. It seems that eq.(\ref{lambda}) is just barely compatible
with our data which give, after taking into account the corrections to scaling,
\be
   \mu = 1.353 \pm 0.003 \;, \qquad g_1 = 0.905 \pm 0.008.
\ee

We also tried to fit $N(t)$ by an exponential plus a constant, the latter 
corresponding to events where the two starting points are on different infinite 
clusters. We obtained $P_{\rm diff} < 10^{-6}$ for $d=2$ and $P_{\rm diff} < 2
\times 10^{-6}$ for $d=3$, both for ${\bf x}-{\bf y} = (1,0)$ and for 
${\bf x}-{\bf y} = (1,1)$.
 
Finally, we should point out that the advantage of our two-seed algorithm over the 
naive one using a single seed is even more pronounced for supercritical percolation. 
There, the survival chance for an event decays exponentially with $t$, while it would 
not decay at all for the growth of a single cluster.

\eject

\end{document}